# Spatiotemporal stabilization proof of concept of Broad Area Semiconductor laser sources

J. Medina[1], W.W. Ahmed[1,2*], S. Kumar[1], M. Botey[1], R. Herrero[1], and K. Staliunas[1,3]

[1]Departament de Física, Universitat Politècnica de Catalunya (UPC), Colom 11, E-08222 Terrassa, Barcelona, Spain
[2]European Laboratory for Non-linear Spectroscopy (LENS), Sesto Fiorentino 50019, Florence, Italy
[3]Institució Catalana de Recerca i Estudis Avançats (ICREA), Passeig Lluís Companys 23, E-08010, Barcelona, Spain
[*]corresponding author: waqas.waseem291@gmail.com

We provide a numerical proof of concept of a stabilization mechanism for BAS laser sources. The scheme is based on the simultaneous introduction of in-phase two-dimensional modulations on the refractive index and pump (gain). We numerically proof total stabilization in BAS laser sources, both in space and time. We also examine the interplay between the index and gain modulations and the effect of the slow relaxation of carriers on the stabilization performance.

Broad area semiconductor (BAS) amplifiers and lasers are promising and reliable light sources, used in many applications due to their compactness and high conversion efficiency. However, a major drawback of such devices is the relatively low spatial and temporal quality of the emitted beam [1,2]. Modulation instability (MI) [3,4] is the fundamental phenomenon that induces transverse mode instabilities, breaking up the mode profile into multiple filaments. Filamentation relies on the dependence of the refraction index on the population inversion [5]. Moreover, strong nonlinear interactions of the optical field with the active media lead to complex spatiotemporal dynamics [6]. Subsequently, the operation of high-power lasers sources is sharply degraded in terms of functionality and fiber coupling. Accordingly, an efficient and compact technique to reduce and control the instability is highly desired.

Recently, an interesting solution was proposed, based on introducing a periodic spatiotemporal modulation of the potential to manipulate and control instabilities in spatially extended nonlinear dynamical systems [7]. The method has shown an improvement in the emission of BAS amplifiers and flat mirror VECSELs [8,9]. Although the role of carrier properties and temporal instabilities were not discussed to stabilize BAS amplifiers.

In this short letter, we consider a full model of BAS laser source including the dynamics of carrier density and carrier diffusion. We study the system dynamics by introducing simultaneous and in-phase two-dimensional (2D) pump and index modulations. In particular, we numerically show the stabilization performance of the proposed mechanism to stabilize BAS laser sources, including the effect of slow relaxation and diffusion of the carrier density on the proposed mechanism.

The paraxial model describing the dynamics of electric field amplitude ($A$) and carrier density ($N$) for the proposed doubly modulated BAS laser source may be expressed as:

$$\frac{\partial A}{\partial z} = \frac{i}{2k_0 n}\frac{\partial^2 A}{\partial x^2} + s[(1-ih)N - (1+\alpha)]A + 4im_2(x,z)k_0 A \qquad (1a)$$

$$\frac{\partial N}{\partial t} = \gamma(N - (N-1)|A|^2 + p + 4m_1(x,z) + D\nabla^2 N), \qquad (1b)$$

where $n$ is the effective refractive index of the semiconductor, $D$ is the carrier diffusion, $\gamma$ is the carriers relaxation rate, $\alpha$ corresponds to losses and $h$ is the Henry factor or linewidth enhancement factor. The $s$ parameter is inversely proportional to the light matter interaction length and $p$ is the pump parameter for Broad Area Semiconductor (BAS) light sources [10]. The in-phase spatial modulations, $m_{1,2}(x,z) = m_{1,2}cos(q_x x)cos(q_z z)$, have the same spatial profile with $q_x$ transverse and $q_z$ longitudinal wavenumbers and correspond to pump and refractive index modulation with amplitudes $m_1$ and $m_2$, respectively. The spatial modulations are assumed in the transverse and longitudinal directions on small spatial scales, i.e. $|q_x| \gg |k_x|$ and $|q_z| \gg |\lambda|$, where $k_x$ and $\lambda$ are the typical transverse wavevector and exponential instability grow parameter. The modulation of the system may be characterized by a geometrical parameter relating both wavenumbers: $Q = 2nk_0 q_z/q_x^2$ where $q_x = 2\pi/d_\perp$, $q_z = 2\pi/d_\parallel$ being $d_\perp$ and $d_\parallel$ denote the transverse and longitudinal period of the modulation, respectively.

In order to perform an experimentally intended simulation, we assume realistic parameters provided in Tab.1. Pump is taken above threshold and carrier relaxation rate, $\gamma$, is normalized to the nonradiative characteristic relaxation time ($\tau_r = 2ns$), with a typical normalized value of 0.01. The transverse and longitudinal wavenumbers of spatial modulations correspond to periods of $d_\perp = 2.2\mu m$ and $d_\parallel = 32\mu m$.

| Width | 140 (μm) |
|---|---|
| Length | 6.4 (mm) |
| Refractive index | 3.3 |
| Carriers diffusion (D) | 0.15 (cm$^2$/s) |
| Linewidth enhancement factor ($h$) | 2.0 |
| Normalized losses ($\alpha$) | 0.1 (μm$^{-1}$) |
| Inverse of light matter interaction length ($s$) | 0.03 (μm$^{-1}$) |
| Normalized carrier relaxation rate ($\gamma$) | 0.01 |
| Pumping intensity ($p$) | 1.2 (A) |
| Pump modulation amplitude ($m_1$) | 0.35 (A) |
| Index modulation amplitude ($m_2$) | 0.001 |
| Transverse modulation wavenumber ($q_x$) | 2.852 (μm$^{-1}$) |
| Longitudinal modulation wavenumber ($q_z$) | 0.1965 (μm$^{-1}$) |

Tab. 1. Realistic values of BAS laser source used in numerical simulations.

In the case of small diffusion, $D \ll 0$, and fast relaxation of carriers, $\gamma \geq 1$, i.e. in the class A laser limit, the carrier density can be adiabatically eliminated from (1b):

$$N = \frac{p + 4m_1(x,z) + |A|^2}{1 + |A|^2}. \tag{2}$$

Direct substitution of (2) into (1a) results in:

$$\frac{\partial A}{\partial z} = \frac{i}{2} \frac{\partial^2 A}{k_0 n \partial x^2} + s\left[\frac{p + 4m_1(x,z) - 1}{1 + |A|^2}(1 - ih) - ih - \alpha\right]A + 4im_2(x,z)k_0 A, \tag{3}$$

Which is the simplified equation studied in other studies of BAS laser sources [10-12].

We perform a numerical study of model Eq.1 using a direct numerical integration method. First, we consider an unmodulated BAS laser source which shows a typical spatial chaotic behavior (see Fig.1a) with the corresponding spectrum (see Fig.1b). The width of the field spectrum indicates the range of the unstable wavenumber, contributing to chaotic dynamics. The additional feature that can be analyzed with the full model is the temporal dynamics of the system. The field shows a highly inhomogeneous distribution in transverse space that becomes unstable in temporal propagation (Fig.1.c-d). We consider three points marked as (a,b,c) in Fig.1c and plotted the evolution of the field in Fig.1e. In all three points, the temporal evolution shows that field stabilizes in time after a transient time of ≈40 *ns* (Fig.1.e).

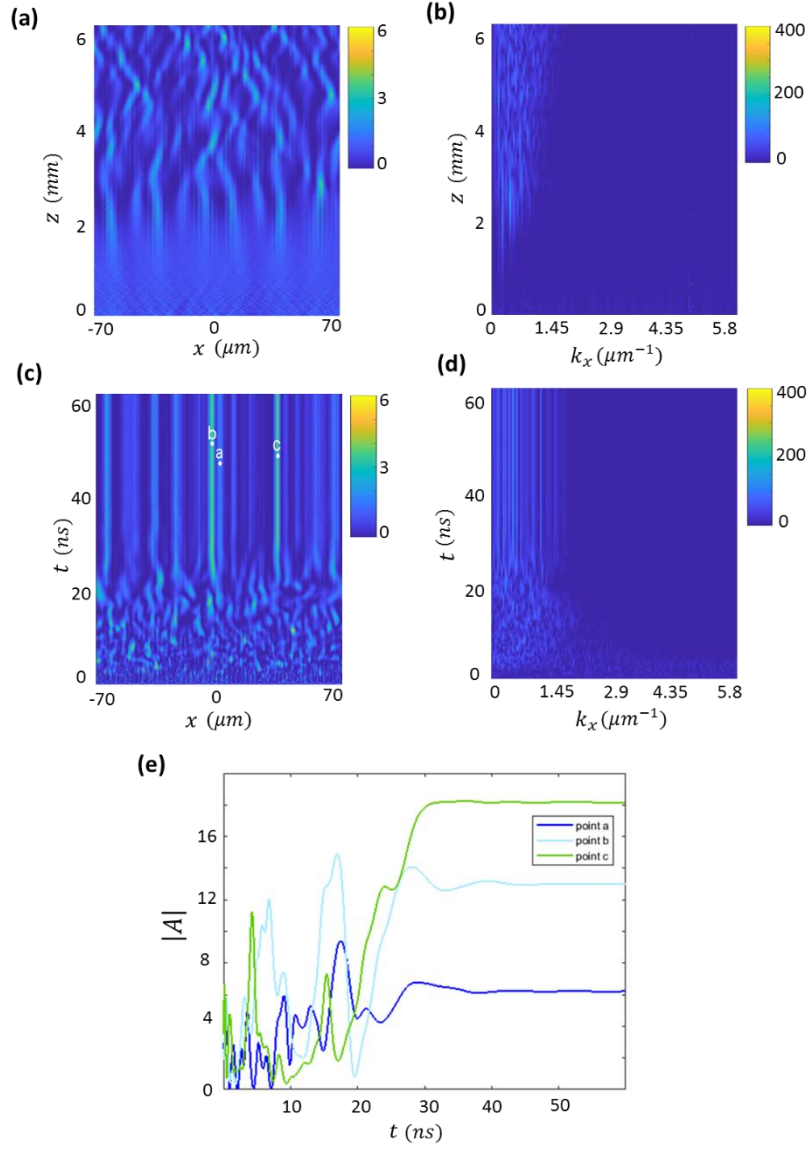

Fig. 1. Direct numerical integration results: for no modulated case $m_1 = m_2 = 0$; (a) Field distribution and (b) Spatial spectrum, $k_x$, inside the semiconductor after a long integration time. Temporal evolution of (c) field and (d) spectrum at the end of the semiconductor. (e) Time evolution at points a,b,c indicated in (c). The simulation parameters are the mentioned on Tab. 1.

Next, we study BAS laser source with only pump modulation and numerical results are shown in Fig.2. In this case, the system is partially stabilized with remaining long wavelength (LW) instabilities as depicted in field spectrum (see Fig. 2b). The inset in field profile (see Fig. 2a) also shows the distortion in the modulated pattern due to LW instability. Moreover, temporal evolution indicates that the transient state becomes shorter in this case, with durations of $\approx 20$ ns. After that, the system shows the oscillations related to relaxational dynamics of Class-B lasers, with characteristic frequencies of the order of the relaxation oscillation frequency, $\omega \approx \sqrt{\gamma}$.

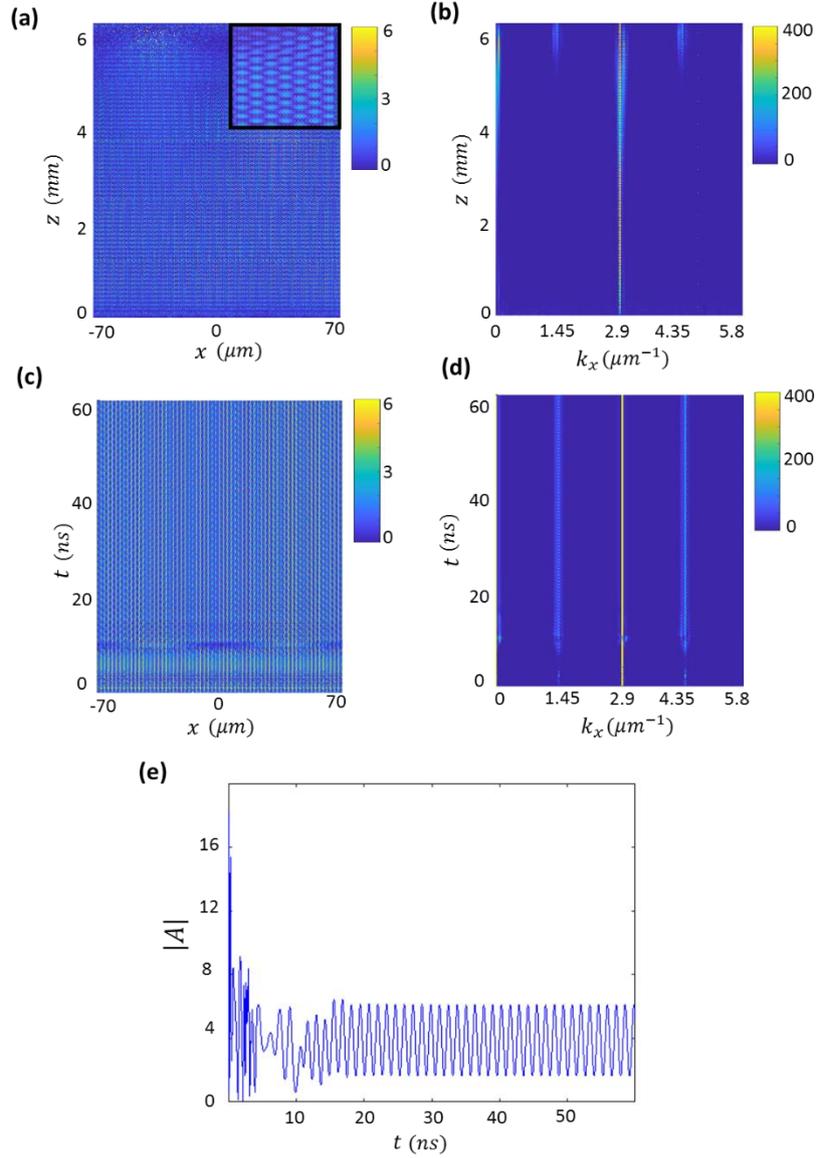

Fig. 2. Direct numerical integration results: for the pump modulated case $m_1$=0.35; $m_2$=0; (a) Field distribution and (b) Spatial spectrum, $k_x$, inside the semiconductor after a long integration time. Temporal evolution of (c) field and (d) spectrum at the end of the semiconductor. (e) Time evolution of one particular point within the semiconductor. The simulation parameters are the mentioned on Tab. 1. The zoomed-in view of the field in modulated case is depicted in the inset.

Finally, we study the doubly-modulated laser source (pump and index modulations) and numerical results are presented in Fig. 3. The result shows that total stabilization is achieved by the simultaneous modulation of pump and refractive index (Fig. 3a). For this case, the spatial spectrum, $k_x$, shows only a component in the transverse modulation wavenumber, $q_x$, demonstrating the total stabilization. This could be also seen in the inset in Fig.3a, where the modulation is perfectly homogeneous.

The transversal spatial distribution in time is totally stabilized under both modulations (Fig.3 c-d). The transient state becomes shorter as compared to the previous case, with a temporal time of ≈ 15 ns. Interestingly introducing the two in-phase modulations, the chaotic spatiotemporal dynamics disappears and the system becomes totally stable in time and homogeneous in space.

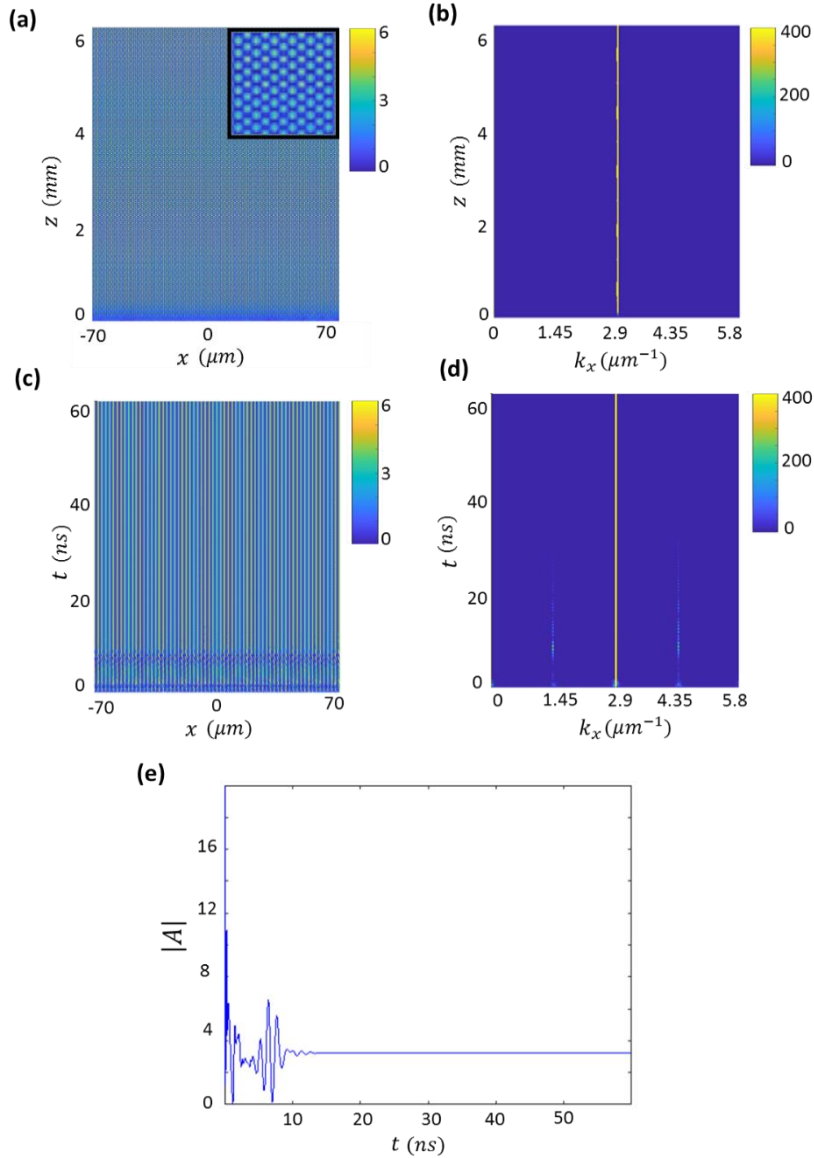

Fig. 3. Direct numerical integration results: for the pump and index modulated case $m_1=0.35$; $m_2=0.001$; (a) Field distribution and (b) Spatial spectrum, $k_x$, inside the semiconductor after a long integration time. Temporal evolution of (c) field and (d) spectrum at the end of the semiconductor. (e) Time evolution of one particular point within the semiconductor. The simulation parameters are the mentioned on Tab. 1. The zoomed-in view of the field in modulated case is depicted in the inset.

---

To conclude, we numerically prove that simultaneous 2D modulations of pump and refractive index provide an efficient scheme to stabilize BAS laser sources, both in space and time. We perform numerical simulations on the full paraxial model considering field and carriers with realistic parameters and the results confirm that doubly-modulated configuration can completely suppress the MI in nonlinear regimes and offer a flexible control on spatiotemporal dynamics of BAS devices.